# Direct Quantification of Water Surface Charge by Phase-Sensitive Second Harmonic Spectroscopy

Laetitia Dalstein, Kuo-Yang Chiang, and Yu-Chieh Wen[*]

Institute of Physics, Academia Sinica, Taipei 11529, Taiwan, R. O. C.

*Correspondence to: ycwen@phys.sinica.edu.tw (Y.C.W.)

## Erratum by the authors

We noticed that the refractive index of the crystalline quartz was not used properly. Correction of this erratum is made in the revised manuscript, which causes changes in values of $\chi^{(2)}$ and $\chi^{(3)}$ in text (as depicted in the resubmitted manuscript with changes remarked) and the scale of y axes in Fig. 2A, 2B, 4B, and Fig. S6, but does not affect any discussion nor conclusion of the manuscript.

## ABSTRACT

We develop and verify a phase-sensitive second harmonic generation spectroscopic scheme that allows for direct determination of the absolute surface charge density and surface potential of a water interface without need of *prior* interfacial information. The method relies on selective probing of surface-field-induced reorientation order of water molecules in the electrical double layer and is, hence, independent of the interfacial molecular bonding structure. Application of this technique to a mixed surfactant monolayer on water suggests the manifest effect of the chain-chain interactions among the monolayer on adsorption of soluble ionic surfactants. We also deduce the third-order nonlinear susceptibility of bulk water and prove its applicability to analysis of charges of various water interfaces. In addition, we show that the Debye-Hückle theory should be avoided in the spectroscopic analysis for its potential significant error, as evidenced experimentally and theoretically.

**TOC GRAPHIC**

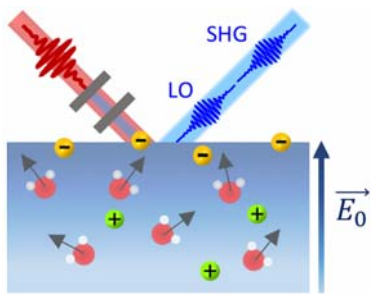

**TEXT**

Electrical double layer (EDL) at water interfaces is an essential ingredient for diverse biological and chemical processes, ranging from protein folding[1-2], atmospheric chemistry[3], to colloidal stability[4-5], and paramount to the development of advanced catalysts[6-8] and energy storage devices[9-10]. Yet, quantitative characterizations of an EDL in atmosphere without model assumption remains a challenge. Electrokinetic measurements yield the electrostatic ($\zeta$) potential at the slipping plane of the EDL, whereas the position of the slipping plane is generally unknown[11]. Potentiometric titration could estimate the surface charge density of the solid/water interface with a particular reaction assumed, but it fails for unknown or multiple origins of the surface charge[12]. Vacuum X-ray photoemission spectroscopy was recently employed to measure the surface potential of nanoparticles[13] and nanoscale liquid films[14], but its applicability could be hampered by the need of synchrotron facilities for high brilliance.

Based on table-top light sources, surface-specific second-harmonic generation (SHG)[15-19] and sum-frequency generation (SFG)[19-27] have been widely used to investigate various water interfaces, but quantification of the surface charge density or the surface potential appeared difficult[18-23]. The problem arises from the long-term inability to separate two signal contributions from a charged interface: One comes from an interfacial monolayer layer where the inversion symmetry is broken by the interface-specific bonding structure (labelled as "the bonded interface layer (BIL)")[15, 24], and the other is induced by water molecules in the EDL with inversion symmetry broken by the surface electric field[15-27]. In many reports, no separate information of the BIL nor EDL can be retrieved from the results[18-23]. In others, one of the two contributions was simply assumed constant or negligible[15, 17, 26-27]. Recently, we demonstrated a SFG scheme for deducing vibrational spectra of the BIL in the presence of an EDL *when* the surface charge density is assumed given or can be measured separately[24]. In another specific situation, Roke *et. al.* showed that BIL and EDL yield different far-field interference patterns for the SHG scattering in dispersed nanoparticle solutions[16], but, whether such difference is discernible for interfaces with smaller curvatures, e.g., a simple planar surface, remains questionable. Very recently, Geiger *et. al.* attempted to interpret phase (and amplitude) changes of the SHG field from the water-silica interface with different ionic strengths in terms of separable EDL and BIL contributions, without knowing absolute phase of the complex effective surface nonlinear susceptibility, $\overleftrightarrow{\chi}^{(2)}_{S,eff}$.[28-29] No justification for validity of the separated EDL or BIL signal, nor quantitative charging property of the EDL could be retrieved from the results.

In this letter, we report the development of a phase-sensitive (PS-) SHG scheme that allows direct characterization of the EDL of charges *selectively*, i.e., without the BIL contribution, for any planar aqueous interface accessible by light. The achievement comes from realization and verification that the imaginary part of $\overleftrightarrow{\chi}^{(2)}_{S,eff}$ for nonresonant SHG is solely contributed from the dc-field-reoriented water molecules in the EDL, through which values of the surface charge density $\sigma$ and surface potential $\phi_0$ can be quantified with the intrinsic third-order nonlinear susceptibility of bulk water, $\overleftrightarrow{\chi}^{(3)}_B$, without need of any *prior* interfacial information. We prove here the measurement principles with the complex $\overleftrightarrow{\chi}^{(3)}_B$ deduced and demonstrate its unique ability in surface chemistry research.

We follow Ref. 26 to formulate SHG process at a charged water interface next to an isotropic medium. The EDL is set up by an atomically thin layer of surface charges at the immediate neighborhood of the interface (at $z \approx 0$, with $z$ along the surface normal) and other ions near the interface by following the Poisson-Boltzmann (PB) distribution[10]. The interfacial structure is affected by the dc-field distribution, $E_0(z)$, and the interfacial bonding interactions in the BIL.

Illumination of the laser field, $\vec{E}(\omega)$, in the medium generates the nonlinear polarization at the SH frequency, $\vec{P}(2\omega)$, as expressed by $\vec{P}(2\omega) = \overleftrightarrow{\chi}^{(2)}:\vec{E}(\omega)\vec{E}(\omega) + \overleftrightarrow{\chi}_B^{(3)}:\vec{E}(\omega)\vec{E}(\omega)\vec{E}_0 + \cdots$.[15, 20, 30] The reflected SHG from such an interfacial system has its field proportional to the effective surface nonlinear polarization $\vec{P}_S^{(2)}$ at $2\omega$, given in terms of $\overleftrightarrow{\chi}_{S,eff}^{(2)}$ by[26]

$$\overleftrightarrow{\chi}_{S,eff}^{(2)} = \overleftrightarrow{\chi}_{BIL}^{(2)} + \overleftrightarrow{\chi}_{EDL}^{(2)}$$

$$\overleftrightarrow{\chi}_{EDL}^{(2)} \equiv \int_0^\infty [\overleftrightarrow{\chi}_B^{(3)}(z') \cdot \hat{z} E_0(z')] e^{i\Delta k_z z'} dz' \cong \overleftrightarrow{\chi}_B^{(3)} \cdot \hat{z}\Psi \qquad (1)$$

with $\Psi \equiv \int_0^\infty E_0(z') e^{i\Delta k_z z'} dz'$,

and $\Delta k_z$ the phase mismatch of reflected SHG. $\overleftrightarrow{\chi}_{BIL}^{(2)}$ denotes the second-order nonlinear susceptibility of the BIL. The effective nonlinear susceptibility of the EDL, $\overleftrightarrow{\chi}_{EDL}^{(2)}$, is uniquely defined here as a consequence of the third-order nonlinear response of water molecules to optical and dc fields[26]. Generally, $\overleftrightarrow{\chi}_{EDL}^{(2)}$ arises mainly from dc-field-induced molecular reorientation[31]. We have used the invariant $\overleftrightarrow{\chi}_B^{(3)}$ of bulk water to approximate the description of $\overleftrightarrow{\chi}_{EDL}^{(2)}$ in Eq. (1) and assumed the electric quadrupole bulk contribution to be negligible[31-32]. In the off-resonant condition, the complex optical susceptibilities of the matter, e.g., $\overleftrightarrow{\chi}_{BIL}^{(2)}$ and $\overleftrightarrow{\chi}_B^{(3)}$, approach real numbers[33], while the effective susceptibilities $\overleftrightarrow{\chi}_{S,eff}^{(2)}$ and $\overleftrightarrow{\chi}_{EDL}^{(2)}$ remain complex because of phase mismatch of the SHG process in a finite-thick EDL. A simple derivation from Eq. (1) then yields

$$\mathrm{Re}\overleftrightarrow{\chi}_{S,eff}^{(2)} = \overleftrightarrow{\chi}_{BIL}^{(2)} + \mathrm{Re}\overleftrightarrow{\chi}_{EDL}^{(2)} = \overleftrightarrow{\chi}_{BIL}^{(2)} + \overleftrightarrow{\chi}_B^{(3)} \cdot \hat{z}\mathrm{Re}\Psi \qquad (2)$$

$$\mathrm{Im}\overleftrightarrow{\chi}^{(2)}_{S,eff} = \mathrm{Im}\overleftrightarrow{\chi}^{(2)}_{EDL} = \overleftrightarrow{\chi}^{(3)}_{B} \cdot \hat{z}\mathrm{Im}\Psi.$$

While $\overleftrightarrow{\chi}^{(2)}_{EDL}$ interferes with $\overleftrightarrow{\chi}^{(2)}_{BIL}$ in the measured $\mathrm{Re}\overleftrightarrow{\chi}^{(2)}_{S,eff}$ and $\left|\overleftrightarrow{\chi}^{(2)}_{S,eff}\right|$, it can now be selectively probed via $\mathrm{Im}\overleftrightarrow{\chi}^{(2)}_{S,eff}$. If $\overleftrightarrow{\chi}^{(3)}_{B}$ of bulk water is known, one can deduce $\mathrm{Im}\Psi$ from the measured $\mathrm{Im}\overleftrightarrow{\chi}^{(2)}_{S,eff} = \mathrm{Im}\overleftrightarrow{\chi}^{(2)}_{EDL}$. With the Gouy-Chapman (GC) theory that solves the PB distribution of ions[10], we can find $E_0(z)$ for a given $\sigma$ and ionic strength, and then use Eq. (1) to calculate $\Psi$ from $E_0(z)$ and the known $\Delta k_z$. Relating the calculated $\mathrm{Im}\Psi$ to the measured value allows us to uniquely find $\sigma$, $E_0(z)$, and the surface potential $\phi_0 \left(= \int_0^\infty E_0(z)dz\right)$. To have an intuition about this application, one could use the Debye-Hückel (DH) theory[10], as an approximation, and Eq. (1) and (2) to derive $\phi_0 = (\mathrm{Im}\chi^{(2)}_{S,eff}/\chi^{(3)}_{B}) \cdot [(\kappa^2 + \Delta k_z^2)/\kappa\Delta k_z]$ with $\kappa$ the invert Debye length ($\kappa = \lambda_D^{-1}$). Clearly, $\phi_0$, and so for $\sigma$, can simply be determined from $\mathrm{Im}\chi^{(2)}_{S,eff}$ without other interfacial information. Note that the DH theory based on a linearized PB distribution is known to break down for high $\phi_0$ ($> \sim 25$ mV)[10]. We shall adopt the GC or a more refined theory below, whereas the validity of the DH model will be examined later.

In our experiments, a nonresonant PS-SHG setup was applied to measure the spectral interferogram created by SHG photons from the sample and a local oscillator (Fig. 1A), yielding the complex $\chi^{(2)}_{S,eff}$ of the sample after normalization against a z-cut quartz reference [see Experimental Section and Supporting Information (SI) section 2 for details]. We adopted a S-in/P-out configuration and presented the results below in MKS units with the Fresnel factors removed, so that our spectroscopic calibration (associated with *zyyz* element of the $\overleftrightarrow{\chi}^{(3)}_{B}$ tensor, $\chi^{(3)}_{B,zyyz}$) can readily be applied to other aqueous interfaces. High measurement precision of this setup was confirmed by measurements of the (charge-neutral) neat water/vapor interface, yielding $|\chi_{S,eff}^{(2)}| = (7.03 \pm 0.50) \times 10^{-23}$ m²/V and the phase of $\chi_{S,eff}^{(2)} = -0.2° \pm 3.5°$. Note that the deduced $\chi_{S,eff}^{(2)}$ is essentially real, as expected for a neutral interface ($\chi_{EDL}^{(2)} = 0$).

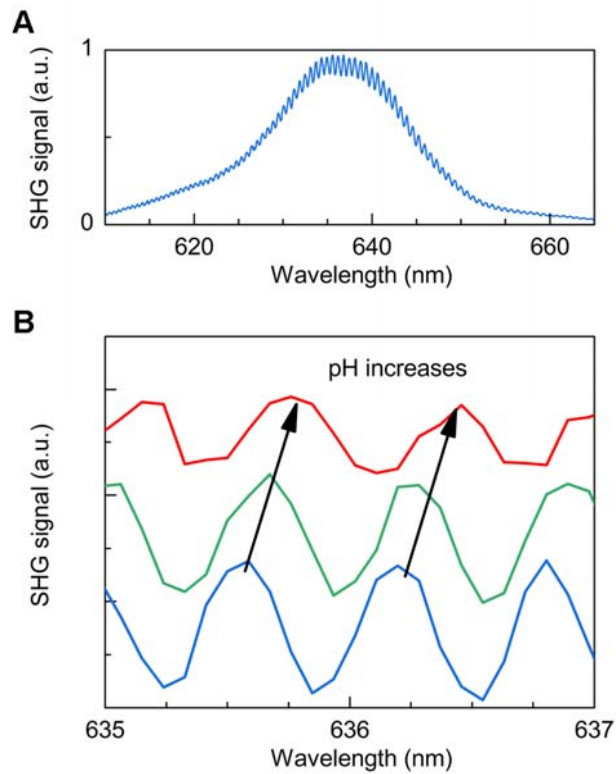

**Figure 1.** SHG interferogram measured by PS-SHG for a lignoceric acid monolayer/water interface at bulk pH = 0.8 [Blue curves in (*A*) and (*B*) for a complete spectrum or zoom in on the fringe], in comparison with pH = 7.6 (green curve) and 10.5 (red curve) in *B*. The spectra in *B* are shifted vertically for clarity.

In inspecting the dc-field effect, the sample studied was a monolayer of lignoceric acid ($C_{23}H_{47}COOH$) on water (see SI section 1 for the sample preparation). Deprotonation of this monolayer (COOH ↔ COO⁻ + H⁺) and the consequent $\sigma$ with pH has been characterized earlier by SFG from the stretch modes of the acid headgroups and the interfacial water[24, 34]; It was shown that the interface is neutral at low pH, but increasingly negatively charged with pH upon deprotonation of the monolayer. Such pH-dependent charging effect can readily be seen in our SHG spectral interferograms of the sample. As shown in Fig. 1B, the phase (and amplitude) of the interferogram varies with pH, revealing the change in phase of $\chi^{(2)}_{S,eff}$ and, thus, its complex nature. It is a direct consequence of the phase mismatch for the nonresonant SHG process in the EDL.

More quantitatively, we analyze the complex $\chi^{(2)}_{S,eff}$ of this sample to verify our PS-SHG scheme and deduce the $\chi^{(3)}_{B,zyyz}$ of bulk water which has not been reported yet. Knowing that the fractional ionization of the monolayer is less than a few percent at pH < 9[24], the signal change of the BIL due to the structural perturbation is hardly detectable[24, 26], so that we can approximate $\chi^{(2)}_{BIL}$ for pH < 9 as that of the neutral interface, denoted by $\chi^{(2)}_{BIL,0}$, but the dc-field-induced $\chi^{(2)}_{EDL}$ of water in the EDL and its change with pH could be already significant, as found in the case of SFG spectra[24, 26, 35]. Therefore, we can obtain $\chi^{(2)}_{EDL}$ from the difference between the measured $\chi^{(2)}_{S,eff} \cong \chi^{(2)}_{BIL,0} + \chi^{(2)}_{EDL}$ and $\chi^{(2)}_{BIL,0}$. At a given pH, we can use the GC theory to find $\sigma$ and $E_0(z)$ from the known surface density of the monolayer and the reported pK$_a$ of the ionization reaction ($\approx 5.25$, confirmed by our previous SFG study[24]; see SI section 3 for the detailed calculation), and then follow Eq. (1) to find $\Psi$ from $E_0(z)$ with the known $\Delta k_z$, and, subsequently, deduce $\chi^{(3)}_B$ from the measured $\chi^{(2)}_{EDL}$ with the calculated $\Psi$.

The measured $\chi^{(2)}_{S,eff}$ for the lignoceric acid monolayer on water versus pH are presented in Fig. 2A, which, again, reveals significant changes in both amplitude and phase. At pH below ~3, $\chi^{(2)}_{S,eff}$ is essentially constant and real within experimental error, revealing that the interface is charge-neutral with $[\chi^{(2)}_{S,eff}]_{\sigma \approx 0} \approx \chi^{(2)}_{BIL,0} = (2.23 - 0.012i) \times 10^{-22}$ m$^2$/V. For 3 < pH < 9, the observed $\chi^{(2)}_{S,eff}$ varies, but with $[\chi^{(2)}_{BIL}]_\sigma \approx \chi^{(2)}_{BIL,0}$, we can deduce $\chi^{(2)}_{EDL}$ from the $\chi^{(2)}_{S,eff}$ and $\chi^{(2)}_{BIL,0}$, and following the GC theory and Eq. (1), find the complex $\chi^{(3)}_B$, which is plotted in Fig. 2B. We expect that if the analysis is correct, the deduced $\chi^{(3)}_B$, being an off-resonant characteristic of the bulk water, should be independent of pH and essentially real. It is indeed what we found. This measurement yields $\chi^{(3)}_{B,zyyz} = (9.56 \pm 1.87) \times 10^{-22}$ m$^2$/V$^2$.

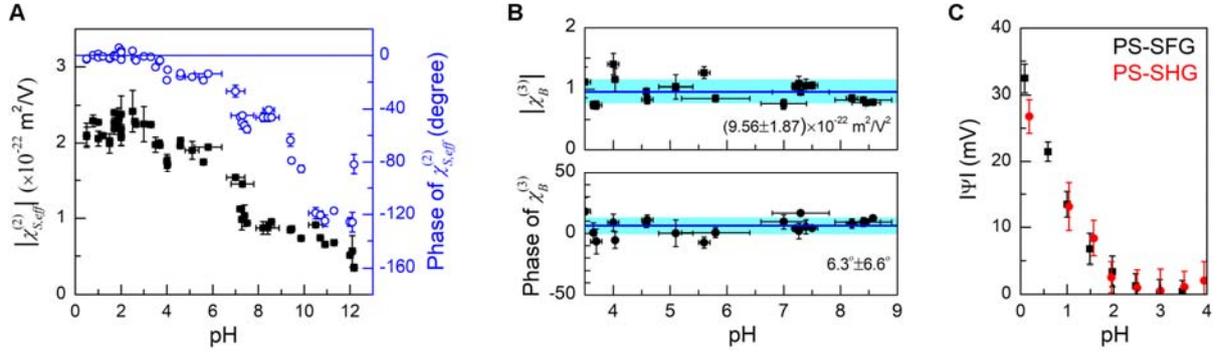

**Figure 2.** Analysis of the PS-SHG results of lignoceric acid monolayer/water and air/water interfaces. (*A*) Amplitude (solid dots) and phase (open dots) of $\chi^{(2)}_{S,eff}$ measured from the lignoceric acid monolayer/water interface at various pH. Line depicts zero in phase. (*B*) Amplitude and phase of $\chi^{(3)}_B$ deduced from the data in *A*, with units of 10$^{-21}$ m$^2$/V$^2$ and degree, respectively. Average and standard deviation of the measured $\chi^{(3)}_B$ are indicated and plotted by horizontal lines with shadowed regions. (*C*) |Ψ| of the air/water interface measured by PS-SHG with different HCl concentrations in water, compared with that obtained from PS-SFG. Good matching of the two spectroscopic results confirms validity of the PS-SHG analysis and that the deduced $\chi^{(3)}_B$ is indeed a bulk water property, independent of the interfacial structure.

We expect $\chi^{(3)}_B$ of bulk water is independent of the interfacial structure and, thus, applicable to analysis of other water interfaces. To confirm the general applicability of our analysis, we use the deduced $\chi^{(3)}_B$ to analyze the air/water interface with ions adsorbed at the surface. Here, the validity of the PS-SHG analysis is justified by its comparison with the PS-SFG spectroscopic analysis that has been proven earlier with the $\chi_B^{(3)}(\omega_{IR})$ SFG spectrum reported[24]. Our comparison is made in terms of the factor $\Psi$ that, for the two spectroscopic setups with identical $\Delta k_z$, should be the same for a given sample. For PS-SHG, we first deduce $[\chi_{S,eff}^{(2)}]_{\sigma=0} = \chi_{BIL,0}^{(2)}$ from the neat water surface. Its value is only ~31 % of that for the lignoceric acid/water interface, indicating distinctly different structures of the two neutral interfaces. With HCl added in water, we, again, approximate $[\chi_{BIL}^{(2)}]_\sigma \approx \chi_{BIL,0}^{(2)}$ for low surface ion coverages[35], and use Eq. (1) to obtain $\chi_{EDL}^{(2)}$ from the difference between the measured $\chi_{S,eff}^{(2)}$ and $\chi_{BIL,0}^{(2)}$. $\Psi$ is then obtained from $\chi_{EDL}^{(2)}$ with the $\chi_B^{(3)}$ deduced earlier via Eq. (1). Similar analyses of $\Psi$ from the $\chi_{S,eff}^{(2)}(\omega_{IR})$ SFG spectra in the OH stretch range were performed with the reported $\chi_B^{(3)}(\omega_{IR})$. (See SI section 4 for details of the SFG analysis). Shown in Fig. 2C are $|\Psi|$ of the air/HCl-solution interface versus pH deduced from the two spectroscopic analyses. Quantitative agreement is found within the experimental accuracy.

With our PS-SHG analysis confirmed, one can now use Eq. (2) and the deduced $\chi_B^{(3)}$ to obtain Im$\Psi$ from Im$\chi_{S,eff}^{(2)}$ for any water interface even with $\chi_{BIL}^{(2)}$ unknown or varying significantly, and then follow the GC theory to determine $\sigma$ or $\phi_0$ from Im$\Psi$. As a demonstration, we investigate adsorption of a prototype ionic surfactant [cetyltrimethylammonium bromide (CTAB)] to the water surface with different surface conditions and bulk concentrations. Shown in Fig. 3A are the surface density of CTA$^+$ (=$\sigma/|e|$ with $e$ the elementary charge) at the water/vapor interface measured by the PS-SHG for different bulk CTAB concentrations, and, in Fig. 3B, the surface density of CTA$^+$ for a 0.2 μM CTAB solution with different surface coverage of insoluble nonionic octadecanol (see SI for the sample preparation). Note that the BIL structure varies upon deposition of the octadecanol[24, 26]. Fig. 3 shows that the surface density of $CTA^+$ increases with its bulk concentration and is obviously enhanced by appearance of the surface octadecanol. The former observation (without octadecanol) agrees with the estimates from the surface tension data[36] (see SI section 5 for the estimation), while the $CTA^+$ density in a mixed surfactant monolayer is directly quantified here for the first time[37]. In interpreting the results, the promotion of the $CTA^+$ adsorption is likely due to the prominent chain-chain interactions between $CTA^+$ and octadecanol in the mixed monolayer, in comparison with the steric effect of molecules and the electrostatic repulsion between the $CTA^+$ headgroups. This finding provides a microscopic implication for explaining the effect of alcohol on the physiochemical properties of micellar solutions[38-39] with importance in many CTAB-related industrial processes[40-41].

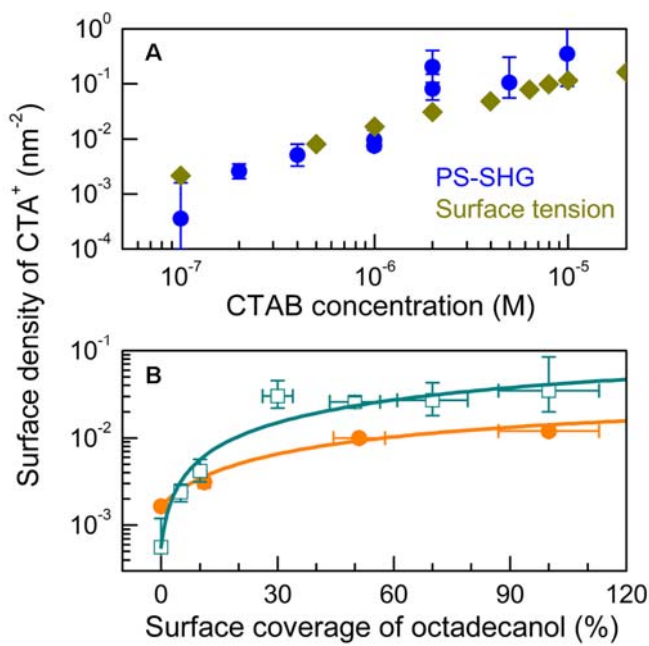

**Figure 3.** Surface density of $CTA^+$ at water interfaces with different surface conditions. (*A*) Results for the water/vapor interface with different bulk CTAB concentrations, deduced from PS-SHG and the surface tension measurements[36]. (*B*) PS-SHG results for the surface of a 0.2 μM CTAB solution with different

surface coverage of insoluble octadecanol molecules. Two sets of data (open and solid dots) are deduced from separated measurements with different batches of CTAB solutions. Lines are guides to the eyes.

To gain further insights into our scheme, we use the GC theory [SI, Eqs. (S7)-(S9)] and Eq. (1) to calculate $\mathrm{Im}\chi^{(2)}_{EDL}$ versus $\sigma$ for different ionic strengths with the known $\Delta k_z^{-1}$. As shown in Fig. 4A, the $\mathrm{Im}\chi^{(2)}_{EDL}$ decreases with an increase in the ionic strength due to a reduced $\lambda_D$, and, interestingly, exhibits a saturation behavior with respect to $\sigma$ for high $\sigma$. This saturation behavior arises from the fact that with high dc-field strengths, the spatial distribution of $E_0(z)$ shrinks by following nonlinear PB distribution[10], for which $\lambda_D$ is no longer a proper characteristic length of the EDL. This effect reduces the phase of $\Psi$ (closer to zero), which compensates the effect of an increasing field strength on $\mathrm{Im}\Psi$ and $\mathrm{Im}\chi^{(2)}_{EDL}$ for high $\sigma$, as depicted in Fig. S5 in SI. To validate this theoretical understanding, we performed an experiment with a monolayer of perfluorotetradecanoic acid ($C_{13}F_{27}COOH$) on NaCl solutions. With a strong tendency to dissociate ($pK_a = -0.3^{42}$), we expect this monolayer to create considerable $\sigma$ that readily saturates the response of $\mathrm{Im}\chi^{(2)}_{EDL}$ to $\sigma$, so that $\mathrm{Im}\chi^{(2)}_{EDL}$ depends simply on the bulk ionic strength. (See SI section 3.2 for an estimation.) Shown in Fig. 4B are the measured $\mathrm{Im}\chi^{(2)}_{S,eff}\left(=\mathrm{Im}\chi^{(2)}_{EDL}\right)$ of the sample for different bulk NaCl concentrations, in comparison to the prediction from the GC theory with the known $\chi^{(3)}_B$ and $\sigma$ set, somewhat arbitrarily, by 5 ~ 100 % of the monolayer dissociation. For such high $\sigma$, the calculated $\mathrm{Im}\chi^{(2)}_{EDL}$ is, again, found insensitive to $\sigma$ but varies with the ionic strength. This calculation agrees with our experiment quantitatively well without adjustable parameter, confirming the theoretical understanding from Fig. 4A. Be aware that the saturation of $\mathrm{Im}\chi^{(2)}_{EDL}$ for high $\sigma$ could in general affect the SFG vibrational spectra in a similar manner[20, 25].

We note that the above understanding (Fig. 4A) remains valid even if the steric effect of ions is taken into account by a refined GC model (see SI section 3 and Fig. S6). On the other hand, the simple DH theory that assumes a linearized PB distribution fails to describe the nonlinear dependence of $\chi^{(2)}_{EDL}$ on $\sigma$ (or $\phi_0$) for high $\sigma$, while this problem is ignored in many SHG/SFG works[16-18, 27-29] with approximated forms of Eq. (1), e.g.,

$$\chi^{(2)}_{EDL} \approx \chi^{(3)}_B \phi_0 \cos\varphi\, e^{i\varphi}, \text{ with } \varphi \equiv \arctan(\Delta k_z \lambda_D). \qquad (3)$$

Shown in Fig. 4A is an example. For an interface with $\sigma = 10^{-2} \sim 10^{-1}$ C/m², e.g., partially ionized oxide surfaces[17] or lipid monolayers on water[25], and an ionic strength of 10 μM, neglecting nonlinearity of the PB distribution by the DH theory [Eq. (3)] could overestimate $\text{Im}\chi^{(2)}_{EDL}$ by 7 ~ 68 times and, hence, yield incorrect interpretation of the result.

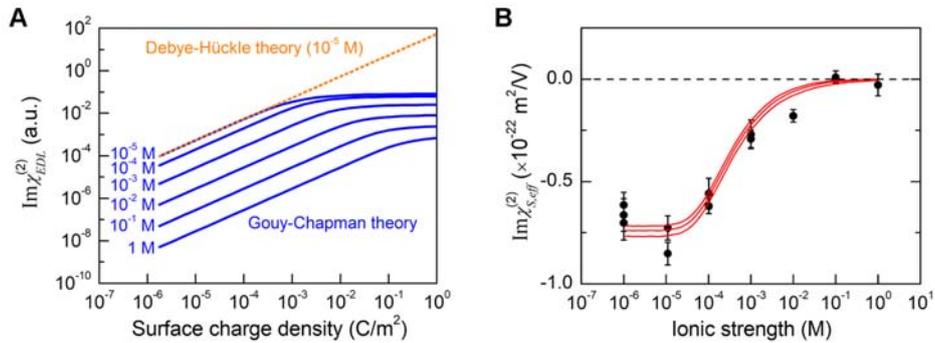

**Figure 4.** Nonlinear dependence of PS-SHG signal on the surface charge density. (*A*) $\text{Im}\chi^{(2)}_{EDL}$ versus surface charge density $\sigma$ calculated by the Gouy-Chapman (GC) theory for different ionic strengths (solid lines). Result from the Debye-Hückel model with an ionic strength of $10^{-5}$ M (dashed line) is shown for comparison. (*B*) Measured $\text{Im}\chi^{(2)}_{S,eff}$ of the perfluorotetradecanoic acid (C₁₃F₂₇COOH) monolayer/water interface with different NaCl concentrations in water (dots). Solid lines are results of the GC theory with $\chi^{(3)}_B$ deduced from Fig. 2B, for $\sigma$ = 0.03, 0.3, and 0.6 C/m² (from top to bottom), corresponding to the ionization fraction of 5, 50, and 100 % of the monolayer. Quantitative agreement confirms the theoretical understanding from *A*.

We have demonstrated and verified a scheme using PS-SHG to selectively probe the surface-field-induced reorientation order of water molecules in the electrical double layer, through which it is now possible to directly find absolute values of the surface charge density and surface potential for various water interfaces without *prior* interfacial information (including spectrum of the bonded interface layer). With the surface charge density known from the PS-SHG, ones can further calculate the SFG spectrum of the EDL, and in turn, find the vibrational spectrum of the bonded interface layer from a PS-SFG measurement[24]. This study will, therefore, offers new opportunities for separate investigations on charging and structural properties of various water interfaces on the microscopic level, leading to advances in our knowledge of interdisciplinary problems, such as erosion and catalytical mechanisms at oxide surfaces, and incorporation of interfacial water onto charged cell membranes.

**EXPERIMENTAL SECTION**

The light source was an optical parametric amplifier pumped by a 1-kHz femtosecond Ti:sapphire laser system (Astrella, Coherent). The fundamental inputs for the PS-SHG setup were 50-fs pulses at 1270 nm. This beam, after passing through a *y*-cut quartz acting as a local oscillator (LO) and a $SrTiO_3$ plate for relative phase modulation, was focused onto the sample surface. The SHG photons generated from the sample in reflection interfered with the LO reflected from the sample and created an SHG spectral interferogram that was measured by a charge-coupled device (CCD)-based polychromator. The interferogram was Fourier-analyzed[22] with normalization against a *z*-cut quartz to yield the phase of $\chi_{S,eff}^{(2)}$, while the amplitude of $\chi_{S,eff}^{(2)}$ was deduced from a separated SHG intensity measurement with the quartz LO removed. The Fresnel coefficients were removed from the presented data (see SI section 2 for details). All measurements were performed with S-in/P-out configuration at room temperature.

To improve precision of the phase measurement, alignment of the sample was set with the help of a white-light interferometer and a quadrant photodetector. We cleaned the *z*-cut reference quartz before measurements and have compared two *z*-cut quartz for excluding possible problems induced by contaminations or the surface polishing. We came to realize that the measurement precision was limited by inhomogeneity of each *z*-cut quartz (~2°) and the long-term stability of the instrument that caused a phase drift (< 0.7°) during each measurement (< 20 min). Accuracy of the phase determination was supported by the measured zero phase angle for several systems, including the neat water surface, LA on >1 mM HCl solutions, and PF on > 0.1 M NaCl solutions (see text for details). Influences of the hyper-Rayleigh scattering (HRS)[19] were excluded in our experiment because (1) incoherent photons generated by HRS cannot interfere the LO, (2) quantitative agreement for the LA/water interface between the measured $\chi_{S,eff}^{(2)}$ and the calculated $\Psi$ versus pH (see main text for the discussion of Fig. 2A and 2B), and (3) quantitative agreement between the PS-SHG and PS-SFG analyses of the HCl-solution/air interface (see main text for the discussion of Fig. 2C).

Note that $\Delta k_z^{-1}$ = 45 nm for SHG from water was essentially the same as that in our PS-SFG setup[26] in which $\Delta k_z^{-1}$ varies between 44 ~ 45 nm in the 3000~3800 cm$^{-1}$.

## Associated Content

**Supporting Information**

Samples preparation, Correction of Fresnel coefficient, Estimation of the surface field near Langmuir monolayers on water, SFG spectroscopic analysis of HCl-solution/air interface, Surface excess of CTAB estimated by the surface tension measurement and Additional figures.

## Author Information

**Corresponding author**

*E-mail: ycwen@phys.sinica.edu.tw

**ORCID**

Laetitia Dalstein: 0000-0002-2591-5421

Kuo-Yang Chiang:

Yu-Chieh Wen:

**Author contributions**

L.D. and Y.C.W. designed the research project. L.D., K.Y.C., and Y.C.W. performed the experiments. L.D. and Y.C.W. discussed the results and wrote the manuscript.

**Notes**

The authors declare that they have no competing interest.

## Acknowledgments

We thank Y. R. Shen for fruitful discussions; J.-R. Huang for help with the surface tension measurement; K. H. Lin for experimental help. L.D. is a recipient of the Academia Sinica Postdoctoral Fellowship. This work was funded by the Ministry of Science and Technology, Taiwan (grant number 106-2112-M-001-001-MY3; 108-2112-M-001-039-MY3).

# References

1. Zhang, L. Y.; Yang, Y.; Kao, Y. T.; Wang, L. J.; Zhong, D. P., Protein hydration dynamics and molecular mechanism of coupled water-protein fluctuations. *Journal of the American Chemical Society* **2009,** *131* (30), 10677-10691.
2. Lo Nostro, P.; Ninham, B. W., Hofmeister phenomena: an update on ion specificity in biology. *Chemical Reviews* **2012,** *112* (4), 2286-2322.
3. Knipping, E. M.; Lakin, M. J.; Foster, F. L.; Jungwirth, P.; Tobias, D. J.; Gerber, R. B.; Dabdub, D.; Finlayson-Pitts, B. J., Experiments and simulations of ion-enhanced interfacial chemistry on aqueous NaCl aerosol. *Science* **2000,** *288*, 301-306.
4. Boroudjerdi, H.; Kim, Y. W.; Naji, A.; Netz, R. R.; Schlagberger, X.; Serr, A., Statics and dynamics of strongly charged soft matter. *Physics Reports-Review Section of Physics Letters* **2005,** *416* (3-4), 129-199.
5. Behrens, S. H.; Borkovec, M., Electrostatic interaction of colloidal surfaces with variable charge. *Journal of Physical Chemistry B* **1999,** *103* (15), 2918-2928.
6. Brown, G. E.; Henrich, V. E.; Casey, W. H.; Clark, D. L.; Eggleston, C.; Felmy, A.; Goodman, D. W.; Gratzel, M.; Maciel, G.; McCarthy, M. I.; Nealson, K. H.; Sverjensky, D. A.; Toney, M. F.; Zachara, J. M., Metal oxide surfaces and their interactions with aqueous solutions and microbial organisms. *Chemical Reviews* **1999,** *99* (1), 77-174.
7. Asahi, R.; Morikawa, T.; Ohwaki, T.; Aoki, K.; Taga, Y., Visible-light photocatalysis in nitrogen-doped titanium oxides. *Science* **2001,** *293* (5528), 269-271.
8. Lopes, P. P.; Strmcnik, D.; Jirkovsky, J. S.; Connell, J. G.; Stamenkovic, V.; Markovic, N., Double layer effects in electrocatalysis: The oxygen reduction reaction and ethanol oxidation reaction on Au(111), Pt(111) and Ir(111) in alkaline media containing Na and Li cations. *Catalysis Today* **2016,** *262*, 41-47.
9. Liu, H.; Logan, B. E., Electricity generation using an air-cathode single chamber microbial fuel cell in the presence and absence of a proton exchange membrane. *Environmental Science & Technology* **2004,** *38* (14), 4040-4046.
10. Bard, A. J.; Faulkner, L. R., *Electrochemical Methods: Fundamentals and Applications*. Wiley: New York, 1980.
11. Delgado, A. V.; Gonzalez-Caballero, F.; Hunter, R. J.; Koopal, L. K.; Lyklema, J., Measurement and interpretation of electrokinetic phenomena. *Journal of Colloid and Interface Science* **2007,** *309* (2), 194-224.
12. Lutzenkirchen, J.; Preocanin, T.; Kovacevic, D.; Tomisic, V.; Lovgren, L.; Kallay, N., Potentiometric titrations as a tool for surface charge determination. *Croatica Chemica Acta* **2012,** *85* (4), 391-417.
13. Brown, M. A.; Abbas, Z.; Kleibert, A.; Green, R. G.; Goel, A.; May, S.; Squires, T. M., Determination of surface potential and electrical double-layer structure at the aqueous electrolyte-nanoparticle interface. *Physical Review X* **2016,** *6* (1), 011007.
14. Favaro, M.; Jeong, B.; Ross, P. N.; Yano, J.; Hussain, Z.; Liu, Z.; Crumlin, E. J., Unravelling the electrochemical double layer by direct probing of the solid/liquid interface. *Nature Communications* **2016,** *7*, 12695.
15. Ong, S. W.; Zhao, X. L.; Eisenthal, K. B., Polarization of water molecules at a charged interface: second harmonic studies of the silica/water interface *Chemical Physics Letters* **1992,** *191* (3-4), 327-335.

16. Gonella, G.; Lutgebaucks, C.; de Beer, A. G. F.; Roke, S., Second harmonic and sum-frequency generation from aqueous interfaces is modulated by interference. *Journal of Physical Chemistry C* **2016,** *120* (17), 9165-9173.
17. Macias-Romero, C.; Nahalka, I.; Okur, H. I.; Roke, S., Optical imaging of surface chemistry and dynamics in confinement. *Science* **2017,** *357* (6353), 784-787.
18. Ohno, P. E.; Saslow, S. A.; Wang, H. F.; Geiger, F. M.; Eisenthal, K. B., Phase-referenced nonlinear spectroscopy of the alpha-quartz/water interface. *Nature Communications* **2016,** *7*, 13587.
19. Dreier, L. B.; Bemhard, C.; Gonella, G.; Backus, E. H. G.; Bonn, M., Surface potential of a planar charged lipid-water interface. what do vibrating plate methods, second harmonic and sum frequency measure? *Journal of Physical Chemistry Letters* **2018,** *9* (19), 5685-5691.
20. Gragson, D. E.; McCarty, B. M.; Richmond, G. L., Ordering of interfacial water molecules at the charged air/water interface observed by vibrational sum frequency generation. *Journal of the American Chemical Society* **1997,** *119* (26), 6144-6152.
21. Tian, C. S.; Shen, Y. R., Structure and charging of hydrophobic material/water interfaces studied by phase-sensitive sum-frequency vibrational spectroscopy. *Proceedings of the National Academy of Sciences of the United States of America* **2009,** *106* (36), 15148-15153.
22. Mondal, J. A.; Nihonyanagi, S.; Yamaguchi, S.; Tahara, T., Three distinct water structures at a zwitterionic lipid/water interface revealed by heterodyne-detected vibrational sum frequency generation. *Journal of the American Chemical Society* **2012,** *134* (18), 7842-7850.
23. Jena, K. C.; Covert, P. A.; Hore, D. K., The effect of salt on the water structure at a charged solid surface: differentiating second- and third-order nonlinear contributions. *Journal of Physical Chemistry Letters* **2011,** *2* (9), 1056-1061.
24. Wen, Y.-C.; Zha, S.; Liu, X.; Yang, S. S.; Guo, P.; Shi, G.; Fang, H.; Shen, Y. R.; Tian, C. S., Unveiling microscopic structures of charged water interfaces by surface-specific vibrational spectroscopy. *Physical Review Letters* **2016,** *116*, 016101.
25. Dreier, L. B.; Nagata, Y.; Lutz, H.; Gonella, G.; Hunger, J.; Backus, E. H. G.; Bonn, M., Saturation of charge-induced water alignment at model membrane surfaces. *Science Advances* **2018,** *4* (3), eaap7415.
26. Wen, Y. C.; Zha, S.; Tian, C. S.; Shen, Y. R., Surface pH and ion affinity at the alcohol-monolayer/water interface studied by sum-frequency spectroscopy. *Journal of Physical Chemistry C* **2016,** *120* (28), 15224-15229.
27. Schaefer, J.; Backus, E. H. G.; Bonn, M., Evidence for auto-catalytic mineral dissolution from surface-specific vibrational spectroscopy. *Nature Communications* **2018,** *9*, 3316.
28. Ohno, P. E.; Chang, H.; Spencer, A. P.; Liu, Y.; Boamah, M. D.; Wang, H. F.; Geiger, F. M., Beyond the Gouy-Chapman model with heterodyne-detected second harmonic generation. *The Journal of Physical Chemistry Letters* **2019,** *10*, 2328-2334.
29. Boamah, M. D.; Ohno, P. E.; Lozier, E.; van Ardenne, J.; Geiger, F. M., Specifics about specific ion adsorption from heterodyne-detected second harmonic generation. *The Journal of Physical Chemistry C* **2019,** *https://doi.org/10.1021/acs.jpcc.9b04425*.
30. Taguchi, D.; Manaka, T.; Iwamoto, M., Imaging of triboelectric charge distribution induced in polyimide film by using optical second-harmonic generation: Electronic charge distribution and dipole alignment. *Applied Physics Letters* **2019,** *114* (23), 233301.
31. Joutsuka, T.; Morita, A., Electrolyte and temperature effects on third-order susceptibility in sum-frequency generation spectroscopy of aqueous salt solutions. *Journal of Physical Chemistry C* **2018,** *122* (21), 11407-11413.

32. Sun, S. M.; Liang, R. D.; Xu, X. F.; Zhu, H. Y.; Shen, Y. R.; Tian, C. S., Phase reference in phase-sensitive sum-frequency vibrational spectroscopy. *Journal of Chemical Physics* **2016,** *144* (24), 244711.
33. Shen, Y. R., *The principles of Nonlinear Optics*. Wiley: New York, 1984.
34. Miranda, P. B.; Du, Q.; Shen, Y. R., Interaction of water with a fatty acid Langmuir film. *Chemical Physics Letters* **1998,** *286* (1-2), 1-8.
35. Pezzotti, S.; Gaigeot, M. P., Spectroscopic BIL-SFG invariance hides the chaotropic effect of protons at the air-water interface. *Atmosphere* **2018,** *9* (10), 396.
36. Zdziennicka, A.; Szymczyk, K.; Krawczyk, J.; Janczuk, B., Activity and thermodynamic parameters of some surfactants adsorption at the water-air interface. *Fluid Phase Equilibria* **2012,** *318*, 25-33.
37. Barnes, G. T.; Lawrie, G. A.; Walker, K., Equilibrium penetration of monolayers. 9. A comparison of treatments for analyzing surface-pressure-area data. *Langmuir* **1998,** *14* (8), 2148-2154.
38. Lu, J. R.; Purcell, I. P.; Lee, E. M.; Simister, E. A.; Thomas, R. K.; Rennie, A. R.; Penfold, J., The composition and structure of sodium dodecyl sulfate-dodecanol mixtures adsorbed at the air-water interface: a neutron reflection study. *Journal of Colloid and Interface Science* **1995,** *174* (2), 441-455.
39. Casson, B. D.; Bain, C. D., Phase transitions in mixed monolayers of cationic surfactants and dodecanol at the air water interface. *Journal of Physical Chemistry B* **1999,** *103* (22), 4678-4686.
40. Laemmli, U. K., Cleavage of structural proteins during the assembly of the head of bacteriophage T4. *Nature* **1970,** *227* (5259), 680-685.
41. Mehta, S. K.; Kumar, S.; Chaudhary, S.; Bhasin, K. K., Effect of cationic surfactant head groups on synthesis, growth and agglomeration behavior of ZnS nanoparticles. *Nanoscale Research Letters* **2009,** *4* (10), 1197-1208.
42. Goss, K. U., The pKa values of PFOA and other highly fluorinated carboxylic acids. *Environmental Science & Technology* **2008,** *42* (2), 456-458.

Supporting Information for

# Direct Quantification of Water Surface Charge by Phase-Sensitive Second Harmonic Spectroscopy

Laetitia Dalstein, Kuo-Yang Chiang, and Yu-Chieh Wen[*]

Institute of Physics, Academia Sinica, Taipei 11529, Taiwan, R. O. C.

*Correspondence to: ycwen@phys.sinica.edu.tw (Y.C.W.)

1. **Samples preparation**

   Lignoceric acid (LA, > 99% purity), perfluorotetradecanoic acid (PF, 96 % purity), NaOH (reagent grade, pellets), HCl (37 wt % water solution, reagent grade), NaCl (reagent, > 99%), cetyltrimethylammonium bromide (CTAB, > 99%), octadecanol (99 %, ReagentPlus), and chloroform (anhydrous grade, stabilized with ethanol) were obtained from Sigma-Aldrich. NaCl was baked at 600 °C for about 6 hours and slowly cooled down to room temperature before use. The other chemicals were used as received. Water was deionized by a Millipore system and had a resistivity of 18.2 MΩ-cm. Its pH and ionic strength were varied by solvation of NaOH, HCl, and NaCl.

   We follow Ref. 1 and 2 to prepare Langmuir monolayers on water. LA, PF, or octadecanol was spread from a chloroform solution onto the aqueous surface, and the solvent was then allowed to evaporate for ~10 minutes. For LA and PF, the monolayer was prepared in a thoroughly cleaned Teflon Langmuir trough by compressing it at a constant rate of 200 µm/s to the surface tension of 6 mN/m, corresponding to the monolayer density of 22 and 27.5 Å$^2$/molecule for LA and PF, respectively. For octadecanol, the monolayer on water was prepared at the equilibrium spreading pressure, with a surface coverage controlled by the given amount of the octadecanol molecules deposited. Different from the Langmuir monolayers, the CTAB monolayers on water were formed by following the Gibbs adsorption kinetics and characterized here by a concentration-dependent surface tension measurement (see *SI*, Fig. S4).

2. **Correction of Fresnel coefficient**

   We follow the spectral analysis described in Ref. 3 to deduce $\chi^{(2)}_{S,eff}$ of the sample in MKS units with the Fresnel coefficients removed. In describing the theory, we generalize SHG as the SFG process with two input fields $\vec{E}(\omega_1)$ and $\vec{E}(\omega_2)$. The reflected SFG from an interfacial system has its field proportional to the effective surface nonlinear polarization $\vec{P}_S^{(2)}$ at the SF frequency $\omega_s (= \omega_1 + \omega_2)$, given by $\vec{P}_S^{(2)}(\omega_s) = \left[\overleftrightarrow{\chi}_{S,eff}^{(2)}(\omega_s; \omega_1, \omega_2)\right]^{w,FC} : \vec{E}(\omega_1)\vec{E}(\omega_2)$, where the superscript "$w, FC$" on $\overleftrightarrow{\chi}_{S,eff}^{(2)}$ denotes the quantity with the Fresnel coefficients included. $\left[\overleftrightarrow{\chi}_{S,eff}^{(2)}\right]^{w,FC}$ is related to $\overleftrightarrow{\chi}_{S,eff}^{(2)}$ without including the Fresnel coefficients, $\left[\overleftrightarrow{\chi}_{S,eff}^{(2)}\right]^{w/o,FC}$, by

$$\left[\overleftrightarrow{\chi}_{S,eff}^{(2)}\right]^{w,FC} = \left[\overleftrightarrow{L}(\omega_s) \cdot \hat{e}_s\right] \cdot \left[\overleftrightarrow{\chi}_{S,eff}^{(2)}\right]^{w/o,FC} : \left[\overleftrightarrow{L}(\omega_1) \cdot \hat{e}_1\right]\left[\overleftrightarrow{L}(\omega_2) \cdot \hat{e}_2\right], \tag{S1}$$

where $\hat{e}_i$ being a unit polarization vector of the optical field at $\omega_i$, and $\overleftrightarrow{L}(\omega_i)$ the tensorial Fresnel factor. For an interface between two continuous media with dielectric constants $\epsilon_1$ and $\epsilon_2$ (with $z$ along the surface normal and the optical plane set on the $x$-$z$ plane), the components of $\overleftrightarrow{L}(\omega_i)$ are expressed as[3]

$$L_{XX}(\omega_i) = \frac{2\epsilon_1(\omega_i)k_{2Z}(\omega_i)}{\epsilon_2(\omega_i)k_{1Z}(\omega_i) + \epsilon_1(\omega_i)k_{2Z}(\omega_i)},$$

$$L_{YY}(\omega_i) = \frac{2k_{1Z}(\omega_i)}{k_{1Z}(\omega_i) + k_{2Z}(\omega_i)},$$

$$L_{ZZ}(\omega_i) = \frac{2\epsilon_1(\omega_i)\epsilon_2(\omega_i)k_{1Z}(\omega_i)}{\epsilon_2(\omega_i)k_{1Z}(\omega_i) + \epsilon_1(\omega_i)k_{2Z}(\omega_i)} \frac{1}{\epsilon'(\omega_i)}, \tag{S2}$$

where $k_{1Z}$ and $k_{2Z}$ are the wavevectors of light projected along the $z$ axis in the two media. For a liquid/air interface ($\epsilon_1 = 1$), we assume the dielectric constant of the interfacial layer, $\epsilon'(\omega_i)$, to be[3]

$$\epsilon'(\omega_i) = \frac{\epsilon_2(\omega_i)[\epsilon_2(\omega_i) + 5]}{4\epsilon_2(\omega_i) + 2}. \tag{S3}$$

One can apply Eq. (S1)-(S3) to SHG with $\omega_1 = \omega_2 = \omega$ and $\omega_s = 2\omega$. Considering a collinear S-in/P-out SHG configuration and the angle of incidence $\theta$, the relation of Eq. (S1) for a liquid sample and the reference $z$-cut quartz can be simplified as

$$\left[\chi_{S,eff}^{(2)}\right]_{sample}^{w,FC} = \sin\theta\, L_{ZZ}(2\omega)\, L_{YY}(\omega)\, L_{YY}(\omega) \left[\chi_{S,eff}^{(2)}\right]_{sample}^{w/o,FC} = F_{sample}\left[\chi_{S,eff}^{(2)}\right]_{sample}^{w/o,FC},$$

$$\left[\chi_{S,eff}^{(2)}\right]_{QZ}^{w,FC} = \cos\theta\, L_{XX}(2\omega)\, L_{YY}(\omega)\, L_{YY}(\omega) \left[\chi_{S,eff}^{(2)}\right]_{QZ}^{w/o,FC} = F_{QZ}\left[\chi_{S,eff}^{(2)}\right]_{QZ}^{w/o,FC}, \tag{S4}$$

where an additional subscript on $\left[\chi_{S,eff}^{(2)}\right]$ refers to the sample or the quartz reference. We have considered the $x$-axis of the $z$-cut quartz crystal set in the incidence plane. SHG from the quartz is essentially contributed from its bulk nonlinearity with $\left[\chi_{S,eff}^{(2)}\right]_{QZ}^{w/o,FC} = i\chi_q l_c$, where $l_c$ is the coherence length of the quartz, and $\chi_q$ the bulk nonlinear coefficient. A simple derivation then relates $\left[\chi_{S,eff}^{(2)}\right]_{sample}^{w/o,FC}$ of our interest to the ratio $\left[\chi_{S,eff}^{(2)}\right]_{sample}^{w,FC}/\left[\chi_{S,eff}^{(2)}\right]_{QZ}^{w,FC}$ derived from a PS-SHG measurement:

$$\left[\chi_{S,eff}^{(2)}\right]_{sample}^{w/o,FC} = i\chi_q l_c F_{QZ} F_{sample}^{-1} \frac{\left[\chi_{S,eff}^{(2)}\right]_{sample}^{w,FC}}{\left[\chi_{S,eff}^{(2)}\right]_{QZ}^{w,FC}} = iA \cdot \frac{\left[\chi_{S,eff}^{(2)}\right]_{sample}^{w,FC}}{\left[\chi_{S,eff}^{(2)}\right]_{QZ}^{w,FC}}$$

with $A \equiv \chi_q l_c F_{QZ} F_{sample}^{-1}$. \tag{S5}

In our SHG experiments, $\theta = 45$ degree, the refractive index $n(2\omega) = 1.332$ and $n(\omega) = 1.323$ for water, $n(2\omega) = 1.54$ and $n(\omega) = 1.53$ for quartz, $l_c = 37.0$ nm, and $\chi_q = 2d_{11}$, where $d_{11}$ refers to the reported nonlinear coefficient for SHG ($d_{11} = 4 \times 10^{-13}$ m/V at 1064 nm) [4]. The factor 2 arises from different conventions in the definitions of $\chi^{(2)}$ and $d$. Eq. (S2)-(S5) then turns out a proportional factor $A = 2.606 \times 10^{-20}$ m²/V in Eq. (S5).

Note that $\chi_q = 4d_{11}$ for SFG due to the number of distinguishable permutations of the input frequencies in our definition of $\vec{P}_S^{(2)}$, in which no degeneracy constant is adopted[5].

3. **Estimation of the surface field near Langmuir monolayers on water**

We follow Ref. 6 and 1 to calculate the ionization fraction of the Langmuir monolayers and the consequent surface field distribution $E(z)$. We adopt the Gouy-Chapman (GC) theory to analyze weakly charged interfaces, while a modified GC model with the steric effect of ions taken in account is also used for highly charged surfaces.

The GC model treats ions as point charges and solves the Poisson-Boltzmann equation for the interfacial potential $\phi(z)$, which for a 1:1 electrolyte solution, takes the form

$$\nabla^2 \phi = 2 \frac{eC}{\epsilon_2} \sinh\left(\frac{e\phi}{k_B T}\right), \tag{S6}$$

with the boundary conditions $\phi(z=0) = \phi_0$ and $\phi = d\phi/dz = 0$ at $z \to \infty$. Here, $C$ is the ionic strength in the bulk subphase, $e$ the elementary charge, $k_B$ the Boltzmann constant, $T$ the temperature, and $\phi_0$ the surface potential. The solution is given by

$$\phi(z) = \frac{4k_B T}{e} \tanh^{-1}\left\{\tanh\left(\frac{e\phi_0}{4k_B T}\right)\exp(-\kappa z)\right\}, \tag{S7}$$

$$\kappa = \left(\frac{2Ce^2}{\epsilon_2 k_B T}\right)^{1/2}, \tag{S8}$$

$$\phi_0 = -\frac{2k_B T}{e}\sinh^{-1}\left\{\frac{\sigma}{(8k_B T \epsilon_2 C)^{\frac{1}{2}}}\right\}, \tag{S9}$$

and the Debye length is $\lambda_D = \kappa^{-1}$, where $\sigma$ is the surface charge density.

For highly charged interfaces, we adopt a modified GC model described in Ref. 7 and 8 to take the finite-size effect of the ions into account. The modified Poisson-Boltzmann equation takes the form

$$\nabla^2 \phi = \frac{eC}{\epsilon_2}\frac{2\sinh\left(\frac{e\phi}{k_B T}\right)}{1 + 2\nu \sinh^2\left(\frac{e\phi}{2k_B T}\right)}, \tag{S10}$$

where $\nu = 2a^3 C$ is the ion packing parameter, and $a$ is the effective size of the counter ions. From the solution of Eq. (S10), we find

$$\phi_0 = -\frac{2k_B T}{e}\sinh^{-1}\left\{\sqrt{\frac{1}{2\nu}\left[\exp\left(\frac{\nu e^2 \sigma^2}{4C\epsilon_2 k_B T}\right) - 1\right]}\right\}. \tag{S11}$$

We assume $a$ to be the Bjerrum length in water at room temperature ($a$ = 0.7 nm) [6], as also adopted by others [7-8].

### 3.1. Lignoceric acid (LA) monolayer on water

We use the GC theory to calculate $E(z)$ for *weakly* ionized LA monolayers at pH < 9. We follow the deprotonation reaction, COOH ↔ COO⁻ + H⁺, to find the surface fraction of COOH and COO⁻, denoted by $X_{COOH}$ and $X_{COO^-}$, with the relations [1, 6]:

$$pK_a = pH_s - \log[(1 - X_{COOH})/X_{COOH}],$$

$$pH_s = pH_b + \frac{e\phi_0}{2.3 k_B T}. \tag{S12}$$

Here, the surface pH (pH$_s$) is related to the bulk pH (pH$_b$) in the above expression through $\phi_0$ which can be obtained from the GC theory via Eq. (S9) in terms of $\sigma$. We also have $\sigma = eN_S X_{COO^-}$ with $N_S$ the surface density of the monolayer. Note that we have neglected the association of COO⁻ with Na⁺ ions from the NaOH solution because of low Na⁺ concentrations at pH below 9.[6]

With pK$_a$ ≈ 5.25 [1, 6, 9-11] and $N_S$ = 4.5 nm⁻², we can deduce $\sigma$, $\phi_0$, and $\phi(z)$ from Eq. (S7)-(S9) and (S12) with the buffering effect of atmospheric CO$_2$ included in the calculation. With $E_0(z)\hat{z} = -\nabla\phi(z)$, we calculate $E_0(z)$ from $\phi(z)$, and then find complex $\Psi$ from Eq. (1) in the main text with $\Delta k_z^{-1}$ = 45 nm, which is plotted as a function of pH in Fig. S1. Note that the calculated $\sigma$ and $\Psi$ lead to a consistent $\chi^{(3)}_{B,zyyz}$ derived from the nonresonant PS-SHG measurements with different pH as explained in the main text, as well as a consistent $\chi^{(2)}_{S,eff}(\omega_{IR})$

SFG spectra in the OH stretch range derived from our previous PS-SFG study with different pH and NaCl concentrations [6], confirming validity of this calculation.

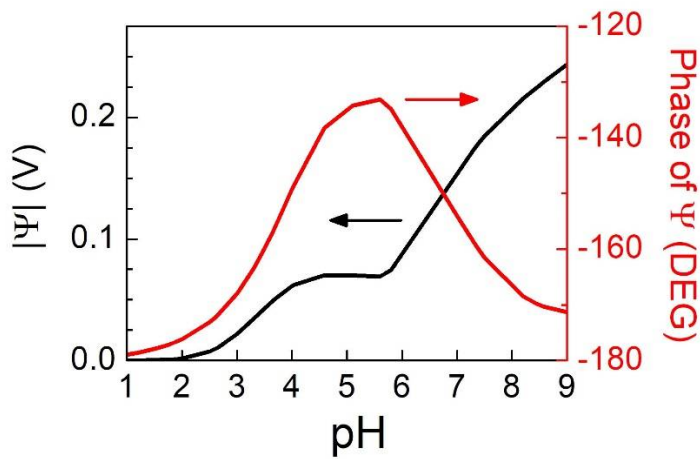

**Fig. S1.** pH-dependent complex $\Psi$ for the LA monolayer/water interface. A plateau in $|\Psi|$ and a maximum in the phase of $\Psi$ at pH 5.6 are caused by an inflection of ionic strength in the solution.

### 3.2. Perfluorotetradecanoic acid (PF) monolayer on water

We use the GC and the modified GC models to estimate the deprotonation fraction of the PF monolayers on water. With the same deprotonation reaction for PF and LA molecules, we can, again, use Eq. (S12) to relate $X_{COO^-}$ (= 1 - $X_{COOH}$) of PF to $pH_s$ and $\phi_0$. With $N_S = 3.6$ nm$^{-2}$ and the literature value of $pK_a$ [$\approx$ –0.3 [12]] for PF molecules, we can, again, use the GC model with Eq. (S9) and (S12) to estimate $X_{COO^-}$, or adopt the modified GC model through Eq. (S11)-(S12). Results from the two models are plotted as a function of $C$ at $pH_b = 6$ in Fig. S2. It is seen that the GC model deviates from the modified GC model for such a highly charged interface due to the steric effect of ions. On the other hand, both models indicate that the monolayer is substantially ionized with $X_{COO^-} > 20\%$, corresponding to $\sigma > 0.12$ C/m², throughout various $C$. Such large $\sigma$ can readily saturate the response of $\text{Im}\chi^{(2)}_{EDL}$ to $\sigma$, as revealed in Fig. 4A in the main text.

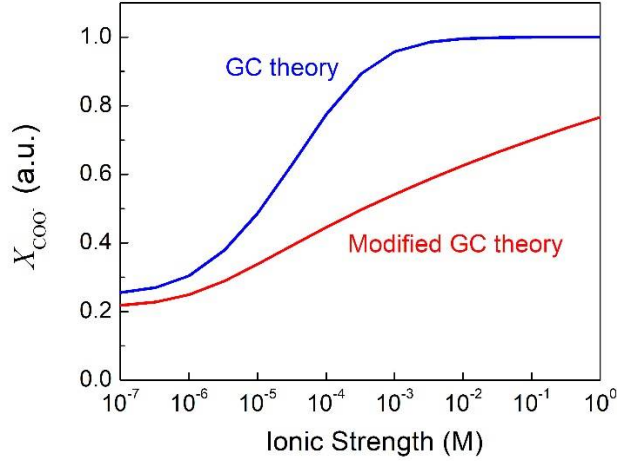

**Fig. S2.** Deprotonation fraction of the perfluorotetradecanoic acid monolayers on water. The calculations are performed with the GC and the modified GC theories with different ionic strengths at $pH_b = 6$.

4. **SFG spectroscopic analysis of HCl-solution/air interface**

We follow Ref. 13 to perform SFG analysis of the dc-field-induced $\Psi$ for the water/air interface. Our broadband PS-SFG spectroscopy has been reported previously[6, 13]. The measured complex $\chi^{(2)}_{S,eff}(\omega_{IR})$ with S-, S-, P- (SSP-) polarized SFG, visible, and mid-IR fields is normalized against the z-cut quartz crystal with the Fresnel coefficients removed. We measured a set of the $\chi^{(2)}_{S,eff}(\omega_{IR})$ SFG spectra of the water/air interface in the OH stretching region with different HCl concentrations in water. Similar to SHG, $\chi^{(2)}_{S,eff}(\omega_{IR})$ obtained from the SFG spectroscopy can be expressed by Eq. (1) in the main text. We obtain $[\chi^{(2)}_{S,eff}(\omega_{IR})]_{\sigma=0} = $ $\chi^{(2)}_{BIL,0}(\omega_{IR})$ from the neat water surface, and approximate $[\chi^{(2)}_{BIL}(\omega_{IR})]_\sigma \approx \chi^{(2)}_{BIL,0}(\omega_{IR})$ for low surface ion coverages. We can then follow Eq. (1) in the main text to deduce complex $\chi^{(2)}_{EDL}(\omega_{IR})$ for different HCl concentrations from the difference between the measured $\chi^{(2)}_{S,eff}(\omega_{IR})$ and $\chi^{(2)}_{BIL,0}(\omega_{IR})$, and further deduce $\Psi$ via $\Psi = \chi^{(2)}_{EDL}(\omega_{IR})/\chi^{(3)}_B(\omega_{IR})$ with the $\chi^{(3)}_B(\omega_{IR})$ spectrum characterized earlier [6]. Fig. S3 shows a representative set of $|\Psi|$ deduced from the spectroscopic SFG data in the bonded-OH spectral range for different HCl concentrations. One would expect that $\Psi$, being a characteristic of the surface field distribution, should be independent of the IR frequency. It is indeed what we found within experimental error, supporting validity of our analysis. Data shown in Fig. 2C in the main text are the average and standard deviation of the data of $|\Psi|$ in the bonded-OH spectral range.

Note that with low $|\Psi|$, we cannot determine the phase of $\Psi$ with meaningful precision, so that only the absolute value of $\Psi$ is examined in this SFG study (Fig. S3 and Fig. 2C in the main text). Details of this spectroscopic measurement and analysis will appear elsewhere.

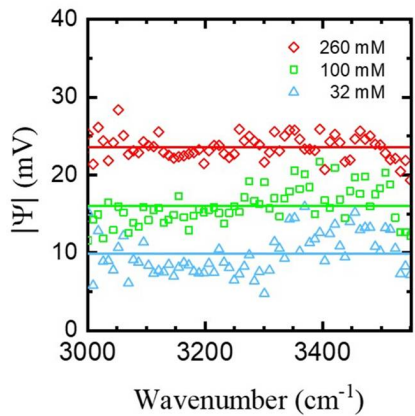

**Fig. S3.** Amplitude of the factor $\Psi$ deduced from the complex $\chi^{(2)}_{S,eff}(\omega_{IR})$ SFG spectra of the air/water interface for different HCl concentrations in water (dots). Lines are the spectral averages.

5. **Surface excess of CTAB estimated by the surface tension measurement**

Adsorption of CTAB molecules has been extensively studied by the equilibrium surface tension, and here we follow Ref. 14 and 15 to examine purity of our sample and determine the surface excess of $CTA^+$ ions. Fig. S4 shows a comparison of the surface tension $\gamma$ versus the bulk CTAB concentration $C_B$ for our sample with a representative set of literature data taken at similar ambient temperature[14]. The results show in general a monotonic reduction in $\gamma$ with $C_B$ until reaching the critical micelle concentration (CMC) of surfactants. Good consistency between the two data (and many other reported results) verifies the quality of our sample.

We followed Ref. 15 to estimate the surface excess, $\Gamma$, of surfactants from $\gamma$ through the Gibbs equation:

$$\Gamma = -\frac{1}{2.303nRT}\frac{d\gamma}{d\log[C_B]}, \quad (S13)$$

where $R$ is the gas constant, and the factor $n = 2$ for uni-univalent ionic surfactants such as CTAB. We have $\sigma = eN_A\Gamma$, where $N_A$ is the Avogadro's number. To compare the estimated $\Gamma$ or $\sigma$ to our PS-SHG results, the precision of the surface tension measurement needs to be better than 0.1 mN/m, which is beyond the limit of our instrument (~0.4 mN/m). Since our sample has $\gamma$ and CMC the same as these reported[14] within the measurement uncertainty, we can use the reported high-precision data of $\gamma$ taken at a similar ambient temperature[14] to estimate $\Gamma$. Shown in Fig. 3A in the main text includes a set of $\Gamma$ data estimated from the reported $\gamma$ [14] through Eq. (S13), which are in reasonable agreement with the PS-SHG results.

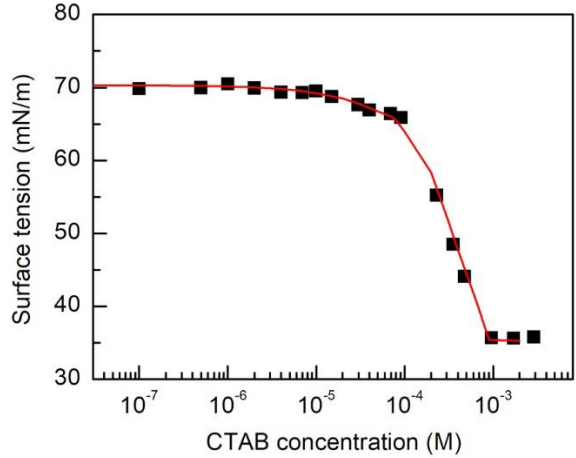

**Fig. S4.** Surface tension measurement of the CTAB solutions. Results for our sample versus the bulk CTAB concentration (black filled square) are shown in comparison with those reported in Ref. 14 (red line). The data from Ref. 14 are vertically shifted by 2.5 mN/m to take the temperature difference into account.

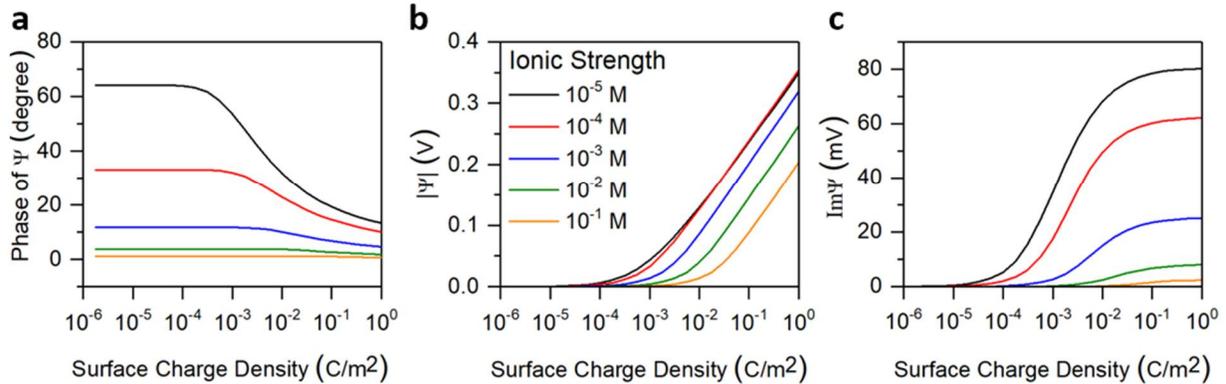

**Fig. S5.** Complex $\Psi$ calculated by the Gouy-Chapman (GC) theory. (a) Phase, (b) amplitude, and (c) imaginary part of $\Psi$ versus the surface charge density for different ionic strengths, calculated by the GC theory through Eq. (S7)-(S9) and Eq. (1) in the main text with $\Delta k_z^{-1} = 45$ nm. It is shown that the plateau of ImΨ for high $\sigma$ is caused by a balance between a decreasing phase angle and an increasing amplitude of Ψ, revealing an intense and more localized $E_0(z)$ at high $\sigma$.

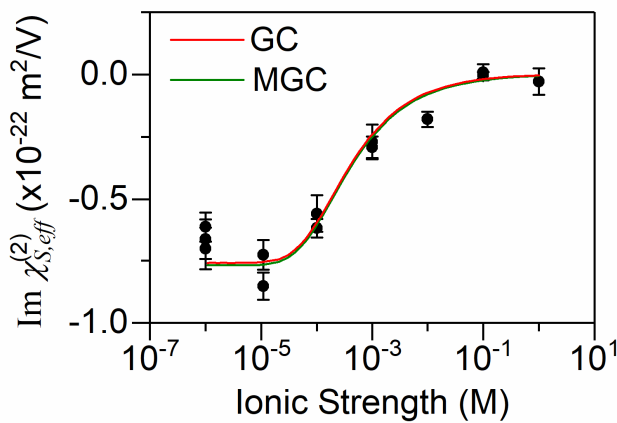

**Fig. S6.** Comparison between the GC and the modified GC theories in terms of $\mathrm{Im}\chi_{EDL}^{(2)}$. Solid lines are $\mathrm{Im}\chi_{EDL}^{(2)} \left(= \mathrm{Im}\chi_{S,eff}^{(2)}\right)$ calculated by the GC theory [Eq. (S7)-(S9)] and the modified GC (MGC) theory [Eq. (S10)-(S11)] with Eq. (1) in the main text, $\Delta k_z^{-1} = 45$ nm, and $\chi_B^{(3)}$ deduced from Fig. 2B in the main text for $\sigma = 0.18$ C/m². The two calculated curves are essentially overlapped, and both agree with $\mathrm{Im}\chi_{S,eff}^{(2)}$ measured from the PF monolayer/water interface with different NaCl concentrations in water (dots). Note that the quantitative agreement does not depend sensitively on $\sigma$ set, as seen in Fig. 4B in the main text.

**Reference**

1. Miranda, P. B.; Du, Q.; Shen, Y. R., Interaction of Water with a Fatty Acid Langmuir Film. *Chemical Physics Letters* **1998**, *286*, 1-8.
2. Tang, C. Y.; Huang, Z. S. A.; Allen, H. C., Binding of Mg2+ and Ca2+ to Palmitic Acid and Deprotonation of the Cooh Headgroup Studied by Vibrational Sum Frequency Generation Spectroscopy. *Journal of Physical Chemistry B* **2010**, *114*, 17068-17076.
3. Wei, X.; Hong, S. C.; Zhuang, X. W.; Goto, T.; Shen, Y. R., Nonlinear Optical Studies of Liquid Crystal Alignment on a Rubbed Polyvinyl Alcohol Surface. *Physical Review E* **2000**, *62*, 5160-5172.

4. Pressley, R. J., *Handbook of Lasers*; Chemical Rubber Co., Cleveland, 1971, p 489, 497.
5. Wei, X., Ph.D. Thesis, University of California at Berkeley (2000).
6. Wen, Y.-C.; Zha, S.; Liu, X.; Yang, S. S.; Guo, P.; Shi, G.; Fang, H.; Shen, Y. R.; Tian, C. S., Unveiling Microscopic Structures of Charged Water Interfaces by Surface-Specific Vibrational Spectroscopy. *Physical Review Letters* **2016**, *116*, 016101.
7. Borukhov, I.; Andelman, D.; Orland, H., Steric Effects in Electrolytes: A Modified Poisson-Boltzmann Equation. *Physical Review Letters* **1997**, *79*, 435-438.
8. Kilic, M. S.; Bazant, M. Z.; Ajdari, A., Steric Effects in the Dynamics of Electrolytes at Large Applied Voltages. I. Double-Layer Charging. *Physical Review E* **2007**, *75*, 021502.
9. Le Calvez, E.; Blaudez, D.; Buffeteau, T.; Desbat, B., Effect of Cations on the Dissociation of Arachidic Acid Monolayers on Water Studied by Polarization-Modulated Infrared Reflection-Absorption Spectroscopy. *Langmuir* **2001**, *17*, 670-674.
10. Usui, S.; Healy, T. W., Zeta Potential of Stearic Acid Monolayer at the Air-Aqueous Solution Interface. *Journal of Colloid and Interface Science* **2002**, *250*, 371-378.
11. Tyrode, E.; Corkery, R., Charging of Carboxylic Acid Monolayers with Monovalent Ions at Low Ionic Strengths: Molecular Insight Revealed by Vibrational Sum Frequency Spectroscopy. *Journal of Physical Chemistry C* **2018**, *122*, 28775-28786.
12. Goss, K. U., The Pka Values of Pfoa and Other Highly Fluorinated Carboxylic Acids. *Environmental Science & Technology* **2008**, *42*, 456-458.
13. Wen, Y. C.; Zha, S.; Tian, C. S.; Shen, Y. R., Surface Ph and Ion Affinity at the Alcohol-Monolayer/Water Interface Studied by Sum-Frequency Spectroscopy. *Journal of Physical Chemistry C* **2016**, *120*, 15224-15229.
14. Zdziennicka, A.; Szymczyk, K.; Krawczyk, J.; Janczuk, B., Activity and Thermodynamic Parameters of Some Surfactants Adsorption at the Water-Air Interface. *Fluid Phase Equilibria* **2012**, *318*, 25-33.
15. Chang, C. H.; Franses, E. I., Adsorption Dynamics of Surfactants at the Air/Water Interface: A Critical Review of Mathematical Models, Data, and Mechanisms. *Colloids and Surfaces a-Physicochemical and Engineering Aspects* **1995**, *100*, 1-45.